\documentclass{PoS}
\usepackage[T1]{fontenc}
\usepackage{cite}
\usepackage{amsmath}
\usepackage{amssymb}
\usepackage{graphicx}

\title{QCD finite T transition \\
Comparison between Wilson and staggered results}

\ShortTitle{QCD finite T transition -- Comparison between Wilson and staggered
results}
\author{Yasumichi Aoki\\
  Brookhaven National Laboratory, Upton NY, USA\\
  E-mail: \email{yaoki@bnl.gov}}

\author{Zolt\'an Fodor\\
  Department of Physics, University of Wuppertal, Germany\\
  Institute for Theoretical Physics, E\"otv\"os University,
  Budapest, Hungary\\ 
  E-mail: \email{fodor@bodri.elte.hu}}

\author{S\'andor D.~Katz\\
  Department of Physics, University of Wuppertal, Germany\\
  Institute for Theoretical Physics, E\"otv\"os University,
  Budapest, Hungary\\ 
  E-mail: \email{katz@bodri.elte.hu}}

\author{K\'alm\'an K.~Szab\'o\\
  Department of Physics, University of Wuppertal, Germany\\
  E-mail: \email{szaboka@general.elte.hu}}

\author{\speaker{B\'alint C.~T\'oth}\\
  Institute for Theoretical Physics, E\"otv\"os University,
  Budapest, Hungary\\ 
  E-mail: \email{tothbalint@szofi.elte.hu}}

\abstract{A quantitative comparison between the finite temperature behaviour
  of the staggered and Wilson fermion formulations are performed. The
  comparison is based on a physical quantity that is expected to be quite
  sensitive to the fermionic features of the action. For that purpose we use
  the height of the peak for $d\chi_s/dT$, where $\chi_s$ is the quark number
  susceptibility.}

\FullConference{The XXV International Symposium on Lattice Field Theory\\
                 July 30 - August 4 2007\\
                 Regensburg, Germany}

\begin{document}

\section{Introduction}

Staggered fermions are computationally faster than Wilson fermions, the
discretization errors scale with ${\cal O}(a^2)$ and due to the well behaving
spectrum of the Dirac operator, light quark masses -- nowadays even the
physical ones -- can be reached.  Note, however, that a straightforward
definition for staggered fermions exists only for 4,8,... quark flavours.  All
works using 2 or $2+1$ flavours of staggered quarks use the fourth root (or
square root) trick to have only one (or two) flavour(s). The action is defined
by taking the fourth root of the fermion determinant. Since such a
prescription is non-local at fixed lattice spacings, it is debated whether it
is equivalent to a local continuum field theory or not (see
e.g.~\cite{Bernard:2006ee,Creutz:2007rk,Kronfeld:2007lat} and references
therein).

Large scale staggered studies are based on the expectation that in the
continuum limit staggered and Wilson results agree.  Clearly, at fixed,
non-vanishing lattice spacings deviations are possible even if the continuum
results are the same. Therefore, any analysis looking for the equivalence or
non-equivalence between the staggered and Wilson formalism should fulfill two
conditions.  First of all, the analysis should be based on several lattice
spacings and a controlled continuum extrapolation should be carried out. At
least three different lattice spacings are needed, for which the asymptotic
scaling behaviour can be already observed (both for staggered and for
nonperturbatively clover-improved Wilson fermions one expects an $a^2$
scaling).  Secondly, the physical quantity chosen as a basis of such
comparison not only has to be well defined and relatively easily measurable,
but also has to be sensitive to the dynamical fermion sector.

It is not so obvious how to find such a sensitive quantity. E.g.~the masses of
different hadrons provide well defined physical quantities, however, these are
not sensitive enough to the dynamical fermion sector. Even the quenched
calculations, where the dynamical fermions are completely omitted, provide the
physical hadron masses with an error about or less then 10\%
\cite{Aoki:2002fd}. Finding measurable differences in the continuum limit of
such quantities would require extremely high precision, thus, prohibitively
large scale calculations.

Thermodynamic observables can be much more sensitive to the fermionic content
of the theory. This fact is related to the singular/non-singular behaviour of
the finite temperature QCD transition.  E.g.~on the one hand the quenched
theory undergoes \cite{Celik:1983wz,Fukugita:1989yb} a first order phase
transition at non-vanishing temperatures (T). For first order phase
transitions different observables behave in a singular way (the latent heat is
infinite or the temperature derivative of the renormalized Polyakov loop has a
discontinuity).  On the other hand the finite temperature transition of QCD
with staggered fermions and physical quark masses has turned out to be a
crossover \cite{Aoki:2006we}. As a consequence, none of the physical
quantities as the function of the temperature are infinite or discontinuous,
at the most they only undergo a rapid change within a narrow temperature
range.  Clearly, the height of such a peak is expected to be quite sensitive
to the details of the fermionic properties of the action.  As we increase or
decrease the quark masses the peak turns out to be more and more singular and
after a while a second order then a first order phase transition region is
reached.  Thus, the fermionic content (quenched or unquenched with physical or
non-physical quark masses) manifests itself in a very pronounced way.

Choosing a physical quantity that undergoes a rapid change and finding the
maximum of its derivative with respect to the temperature may provide a
quantity sensitive to the finite temperature behaviour of the system.

The aim of this study is to provide a quantitative comparison between the
behaviour of the staggered and Wilson fermion formulations.  We attempt to
perform an analysis at two different lattice spacings (thus no conclusive
continuum extrapolation can be done yet). The observable we have chosen is
related to the transition at non-vanishing temperatures and expected to be
sensitive to the fermionic properties of the theory.

One physical quantity that undergoes a rapid change around the transition
temperature is the quark number susceptibility, which is defined via
\cite{Bernard:2004je}
\begin{equation}
    \left. \frac{\chi_s}{T^2} = \frac{1}{TV} \, \frac{\partial^2 \log
      Z}{\partial \mu^2} \right|_{\mu=0},
\end{equation}
where $\mu$ is the quark chemical potential. The quark number susceptibility
can be directly measured, and it automatically has the correct continuum limit
without the need for renormalization. In addition, the maximum of its
derivative with respect to the temperature, that is, the rate at which the
susceptibility changes during the transition, is sensitive to the dynamical
fermion sector. These properties make the quark number susceptibility a good
candidate for the quantity that a comparison of the Wilson and staggered
fermion formulations should be based on.  Since the rooting procedure of the
determinant is less transparent (might be more problematic) for odd number of
flavours we use three flavours.  (The one flavour theory, where there is no
chiral symmetry breaking \cite{Creutz:2006ts}, is not suitable for our
purposes.)

\section{Action parameters}

The three flavours were degenerate for both Wilson and staggered calculations,
and the gauge action used was the Symanzik tree-level improved gauge action
\cite{Symanzik:1983dc}. In both cases the lattices sizes were $32^3\times 8$
and $32^3\times 10$, and the configurations were generated using the Rational
Hybrid Monte-Carlo algorithm \cite{Clark:2003na}.

\paragraph{Wilson calculations:}
Three steps of stout smearing \cite{Morningstar:2003gk} with smearing
parameter $\varrho=0.1$ were used. The gauge coupling constant was in the
range $\beta = 3.2 - 3.7$.  In addition the femionic sector was clover
improved \cite{Sheikholeslami:1985ij} with a tree level clover coefficient
$c=1.0$.  Note, that for this type of smeared fermions the tree level clover
coefficient essentially leads to an ${\cal O}(a)$ improved action
\cite{Hoffmann:2007nm}.

\paragraph{Staggered calculations:}
Two steps of stout smearing with smearing parameter $\varrho=0.15$ was used
\cite{Aoki:2005vt}, and the gauge coupling constant was in the range $\beta =
3.5 - 4.0$.  This staggered action within this lattice spacing range was shown
to be in the scaling regime \cite{Aoki:2006br}.

\section{Setting the scale}

To be able to make sure that the staggered and Wilson calculations are
performed at the same set of physical parameters, the line of constant physics
was defined by fixing the ratio of the pseudoscalar and the vector meson
masses $m_{\text{PS}}/m_{\text{V}}$.

If the finite temperature behaviour of the Wilson and the staggered
formulations are different, the difference is likely to be most apparent at
small quark mass parameters.  For small quark masses we are closer and closer
to the first order phase transition region, thus all the differences are
easier to see in the height of the peak of our observable.  Decreasing the
quark mass, however, causes the computational costs to rise. Therefore, one
needs to find a compromise between making the comparison more sensitive by
lowering the quark mass (or in other words lowering
$m_{\text{PS}}/m_{\text{V}}$) and keeping the computational costs reasonable.

\begin{figure}
\begin{center}
\resizebox{!}{49.81mm}{\includegraphics{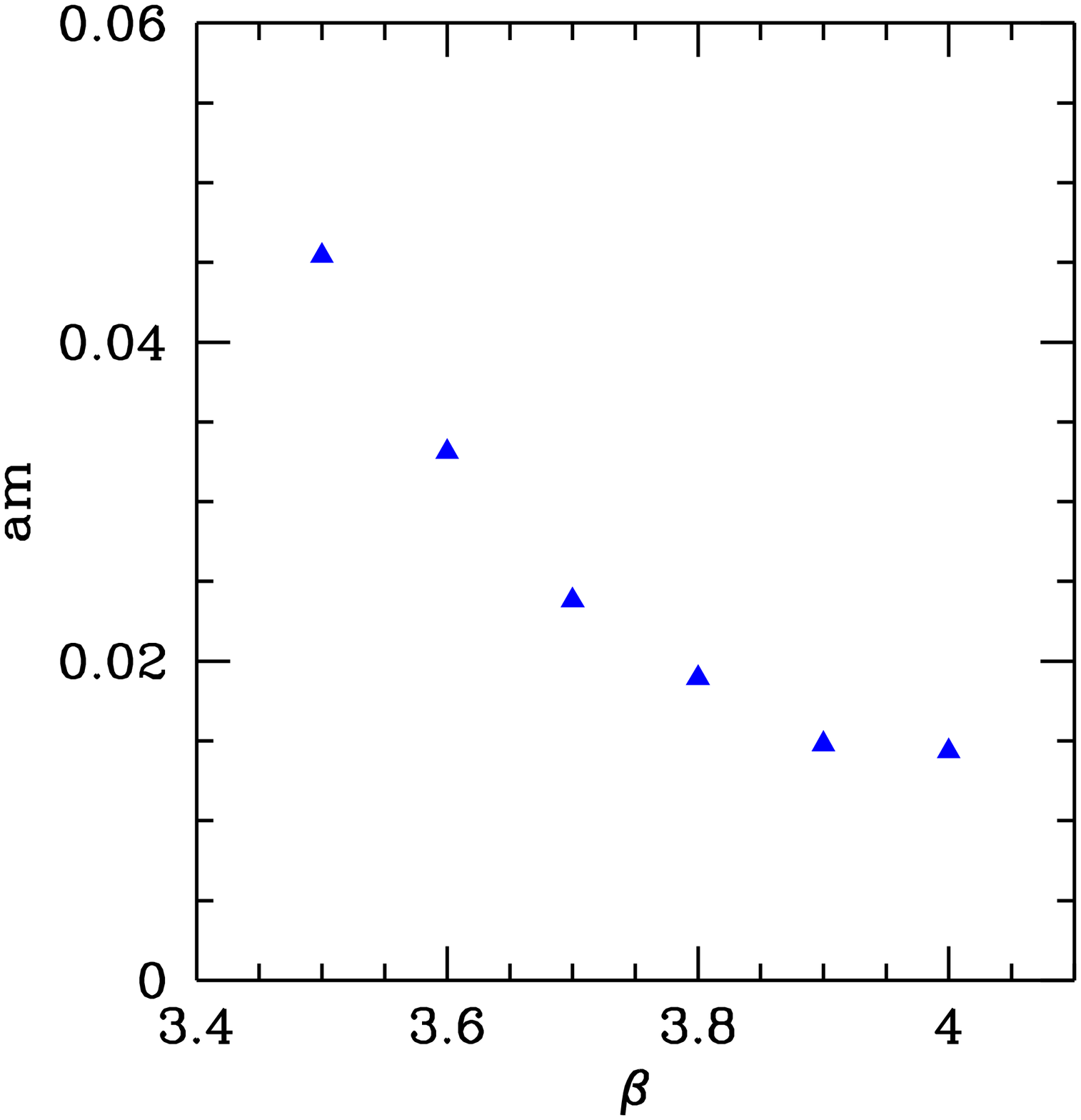}}
\resizebox{!}{49.81mm}{\includegraphics{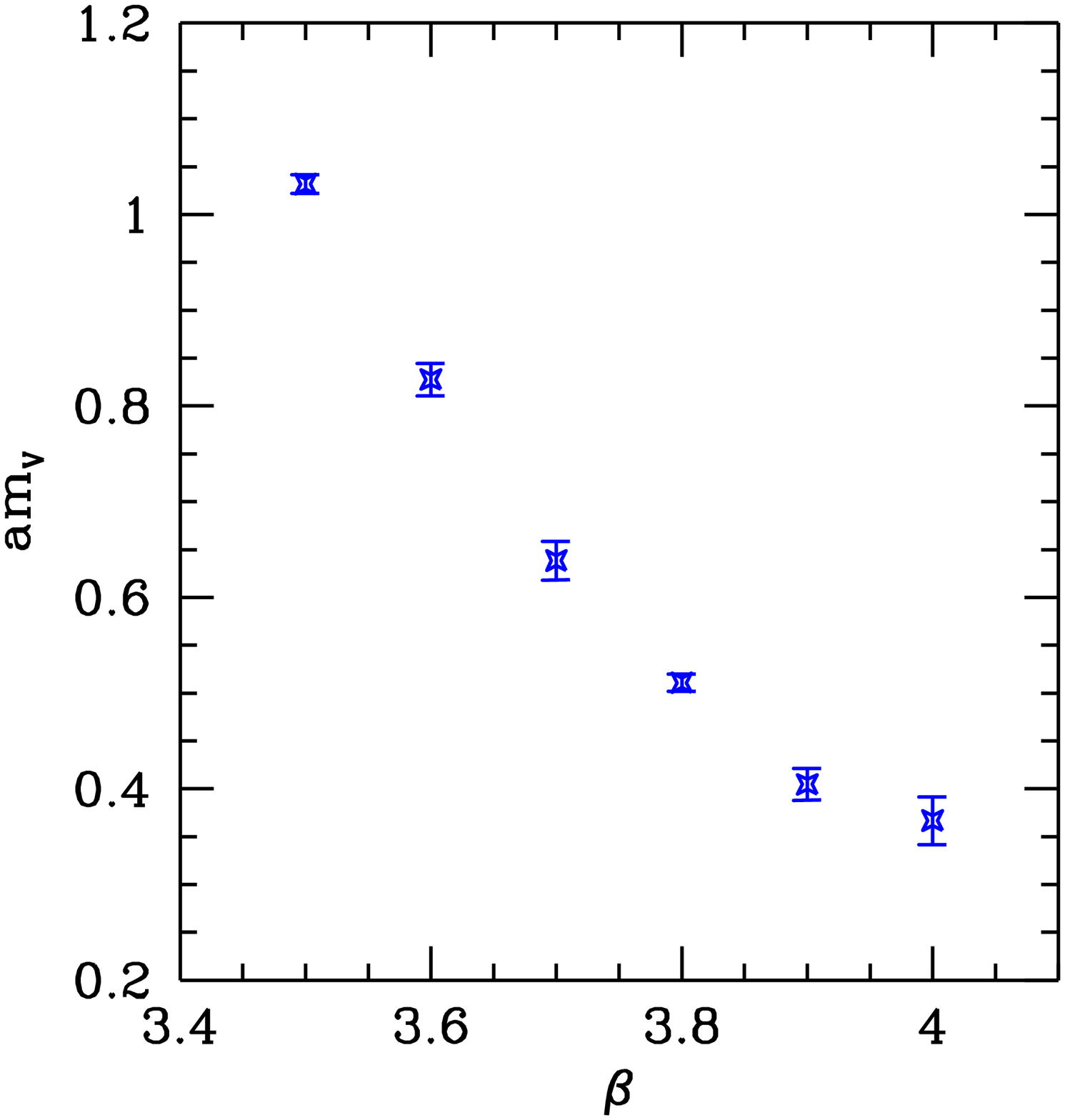}}
\end{center}
\caption{The scale for the staggered calculations: the bare staggered quark
  mass (left panel) and the vector meson mass (right panel) as a function of
  the gauge coupling constant.}
\label{fig:stag_scale}
\end{figure}

As a compromise we have chosen $m_{\text{PS}}/m_{\text{V}} = 0.5$, which sets
the quark mass about $m_s/3$, where $m_s$ is the physical strange quark mass.
This relationship completely defines the line of constant physics. In three
flavour lattice QCD we have two parameters. One of them is the quark mass
which is essentially set by the relationship between the pseudoscalar and
vector mass ratio. The other one is the lattice spacing, which is dominantly
given by the gauge coupling.

\begin{figure}
\begin{center}
\resizebox{!}{49.81mm}{\includegraphics{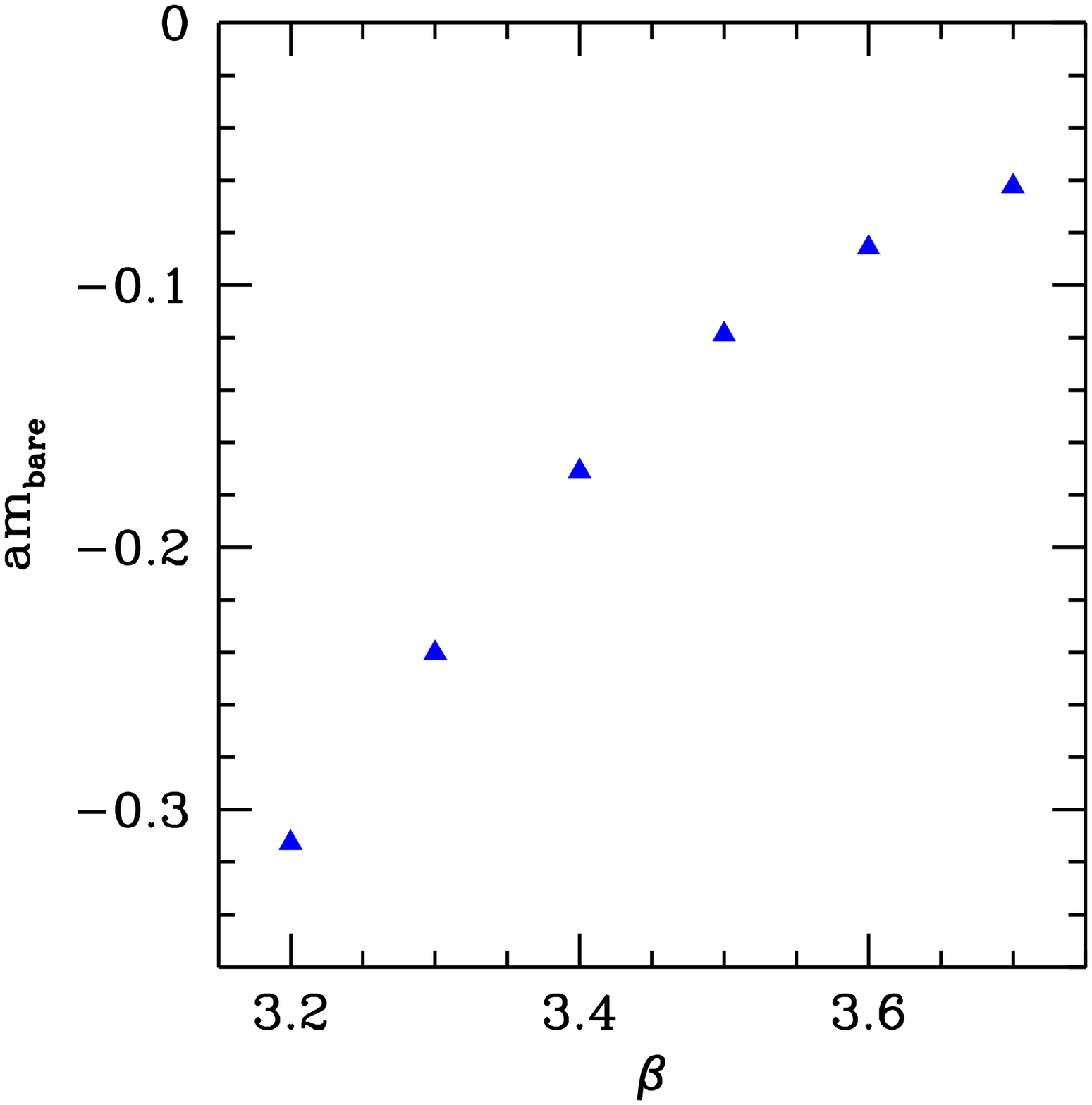}}
\resizebox{!}{49.81mm}{\includegraphics{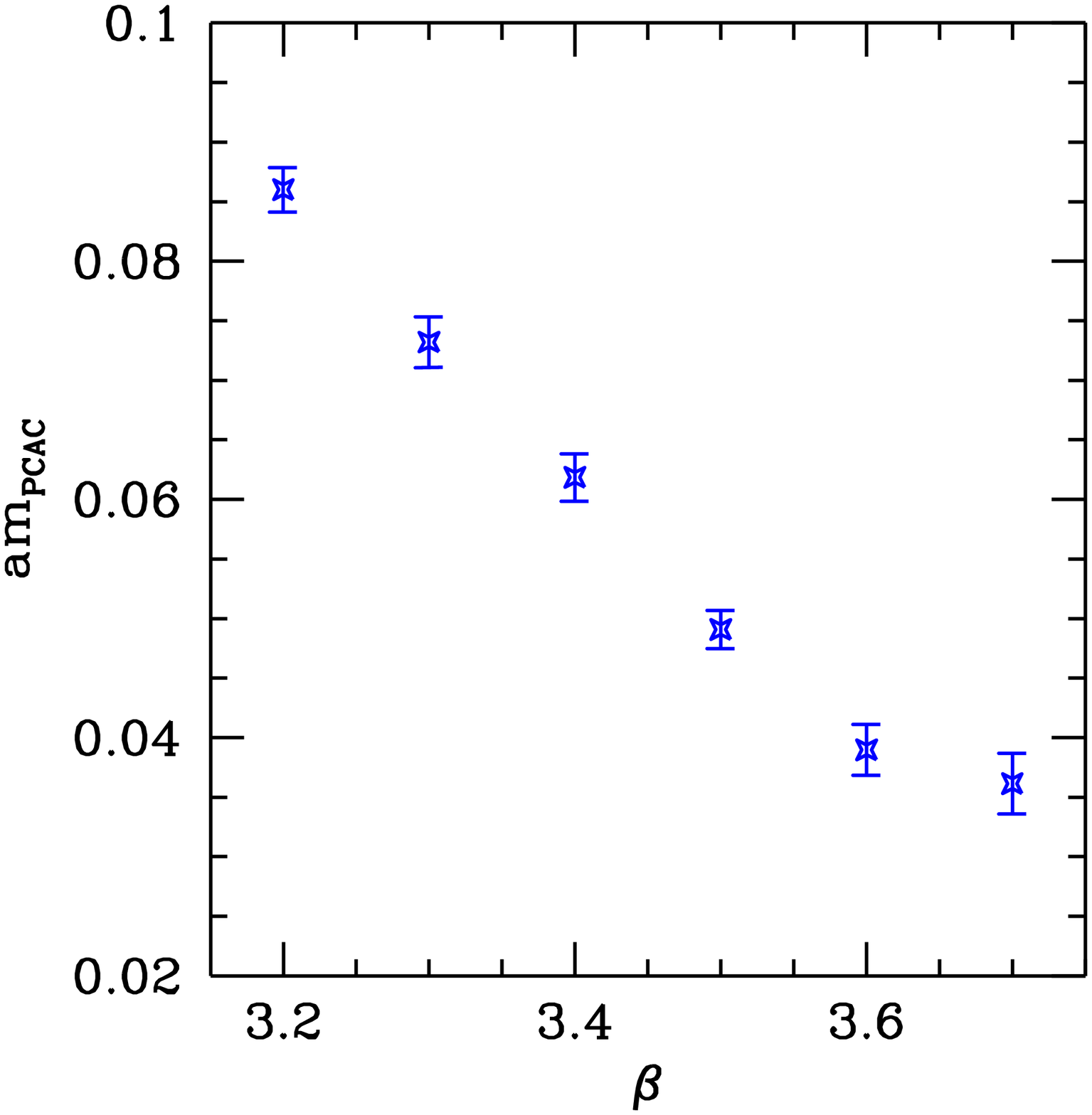}}
\resizebox{!}{49.81mm}{\includegraphics{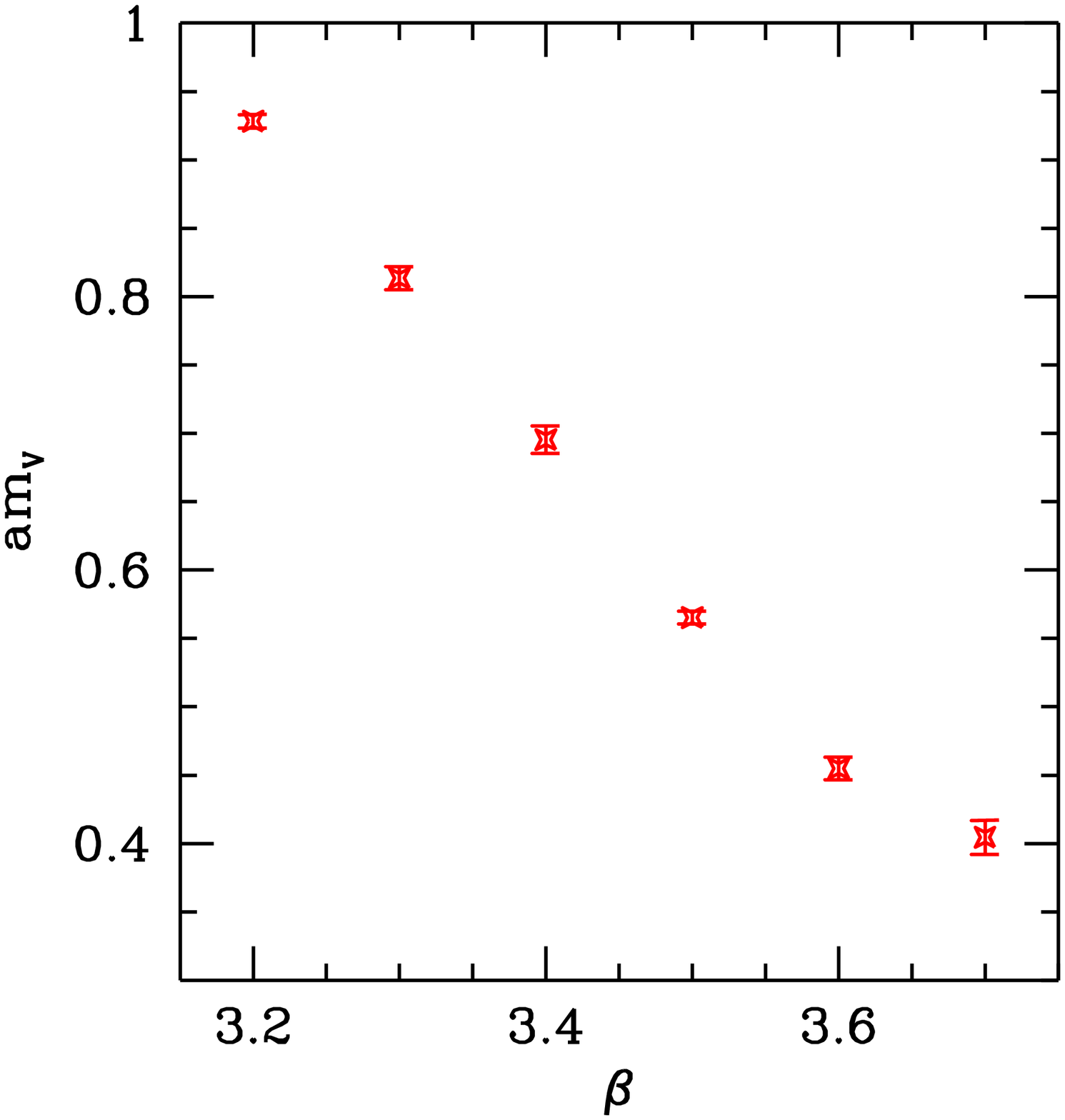}}
\end{center}
\caption{The scale for the Wilson calculations: the bare Wilson quark mass
  (left panel), the PCAC quark mass (middle panel) and the vector meson mass
  (right panel) as a function of the gauge coupling constant.}
\label{fig:wil_scale}
\end{figure}

The bare quark mass and the vector meson mass corresponding to the different
gauge couplig values for the staggered case are shown in Figure
\ref{fig:stag_scale}. For the Wilson calculations, the bare quark mass, the
current algebra quark mass and the vector meson mass are shown in Figure
\ref{fig:wil_scale}.

\section{Results}

\begin{figure}
\begin{center}
\resizebox{!}{70mm}{\includegraphics{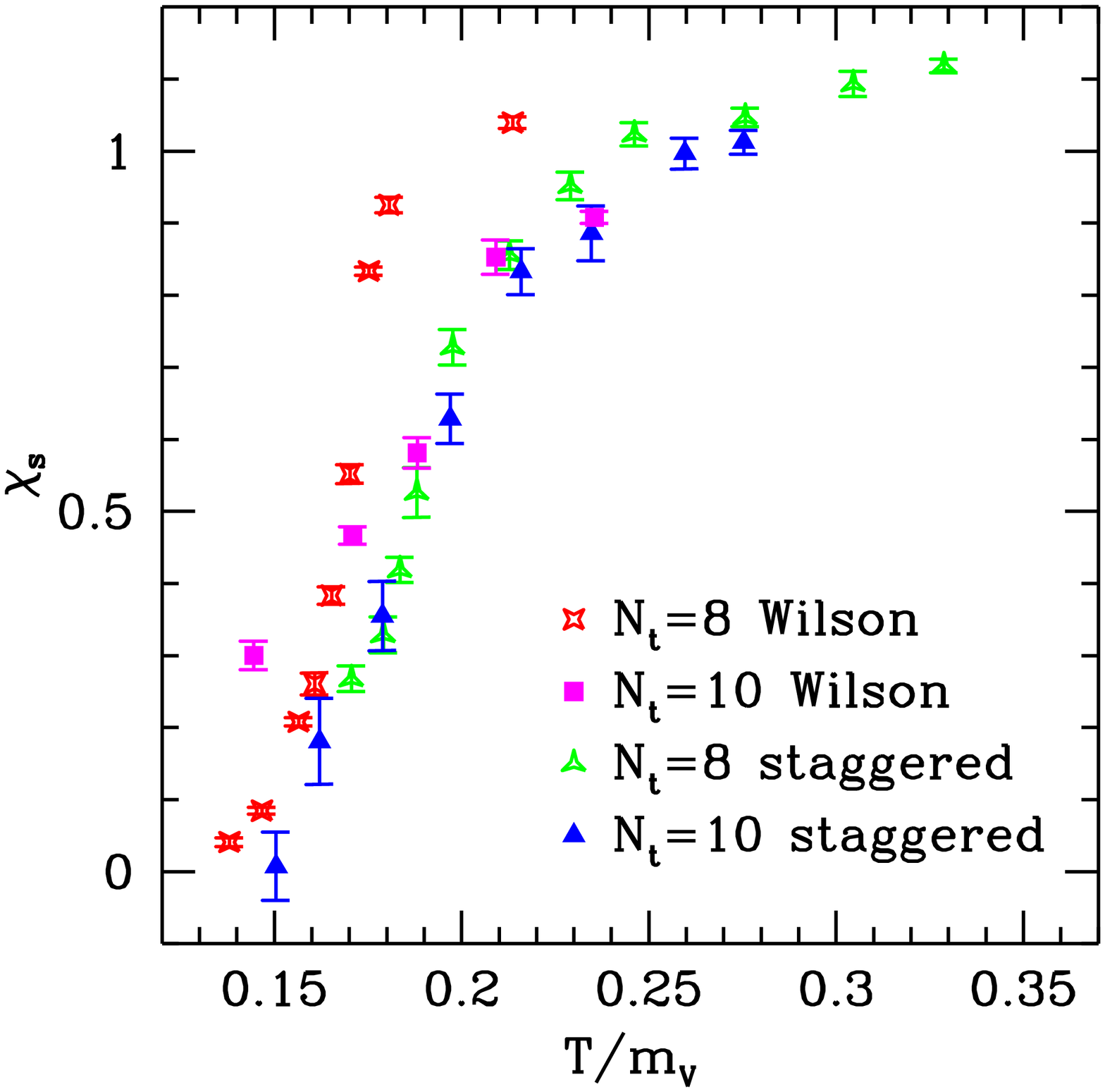}}
\resizebox{!}{70mm}{\includegraphics{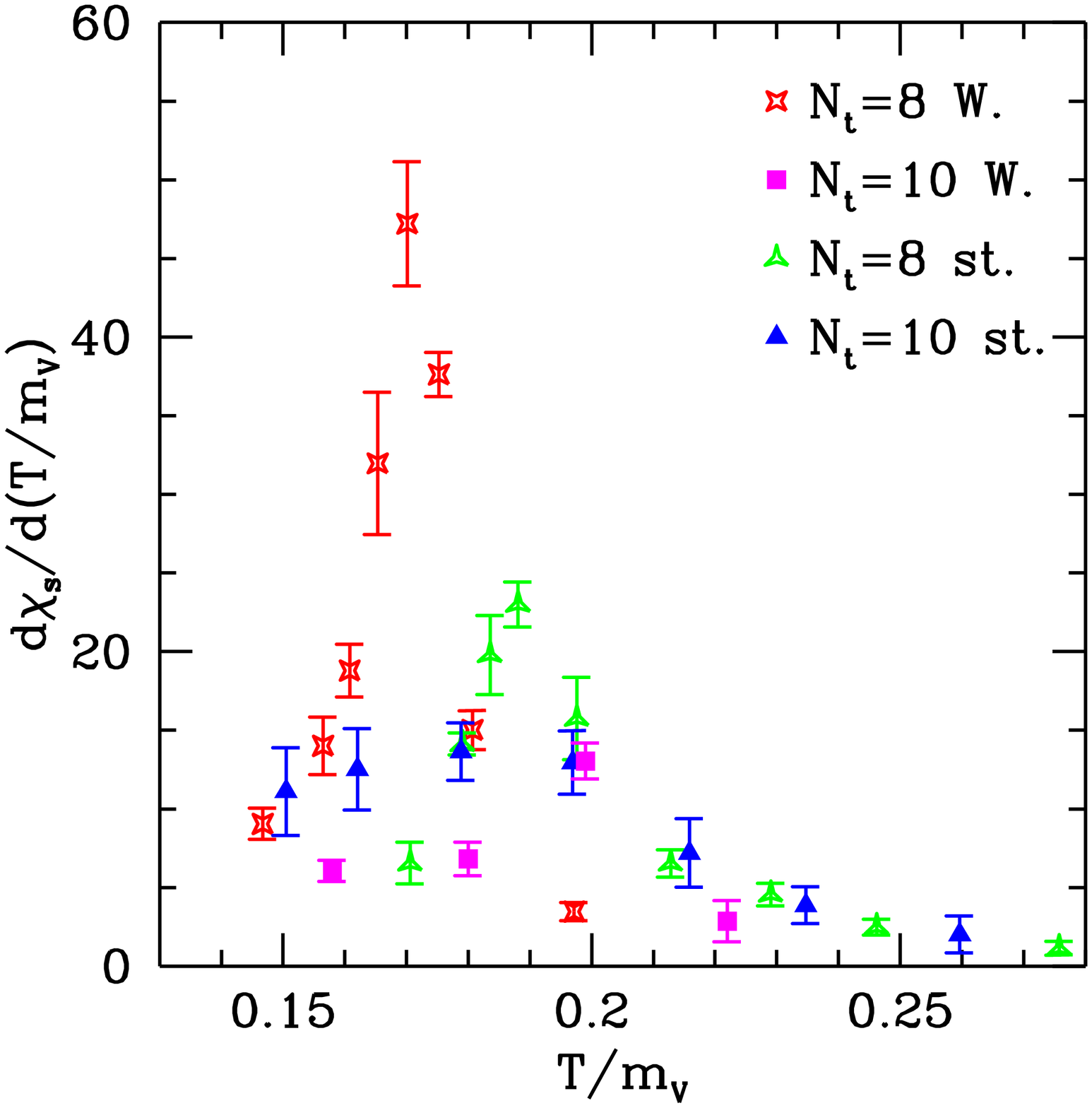}}
\end{center}
\caption{The quark number susceptibility (left panel) and its derivative
  (right panel) as a function of the temperature normalized with the vector
  meson mass.}
\label{fig:qnsusc}
\end{figure}

To be able to compare the staggered and Wilson results the temperature was
made dimensionless by dividing by the vector meson mass. The quark number
susceptibilities as a function of $T/m_{\text{V}}$ are shown in the left panel
of Figure \ref{fig:qnsusc}. The derivative of the susceptibility was obtained
by fitting cubic polynomials to the susceptibility points, then taking the
derivative of the polynomial. The slight change due to the variation of the
fitting range is taken as a systematic error. These derivatives are shown in
the right panel of Figure \ref{fig:qnsusc}.

\section{Conclusions}

Calculating the maximum of the derivative of the quark number susceptibility
with respect to the temperature yields a sensitive quantity, based on which
the finite temperature behaviour of the Wilson and the staggered fermion
formulations can be compared. We have performed calculations using lattices
with temporal extensions $N_t=8$ and $10$. At $N_t=8$ the maximum of the
dervative in the Wilson case is a factor of 2 higher than the staggered
result, whereas at $N_t=10$ the heights of the peaks get closer to one
another. In order to be conclusive one needs to get one or more steps closer
to the continuum limit by performing calculations using $N_t=12$ or even finer
lattices.

\section*{Acknowledgements}

This research was partially supported by OTKA Hungarian Science Grants
No.\ T34980, T37615, M37071, T032501, AT049652, EU Grant I3HP and by DFG
German Research Grant No.\ FO 502/1-1.  The computations were carried out on
the 370 processor PC cluster of E\"otv\"os University, on the 1024 processor
PC cluster of Wuppertal University and on the Blue Gene/L in Research Centre
J\"ulich. We used a modified version of the publicly available MILC code
\cite{MilcCode} with next-neighbor communication architecture for PC-clusters
\cite{Fodor:2002zi}.

\bibliographystyle{pos_lat2007_232}
\bibliography{pos_lat2007_232}

\end{document}